\documentclass[12pt]{iopart}

\usepackage{graphicx}
\usepackage{iopams}
\usepackage[latin1]{inputenc}
\usepackage{amssymb}

\begin{document}

\title[Andreev reflection in WC$_x$ nanowires point contacts]{Andreev reflection measurements in nanostructures
with amorphous WC$_x$ superconducting contacts }


\author{J. Barzola-Quiquia$^1$, M. Ziese$^1$, P. Esquinazi$^1$ and N. Garc\'ia$^{1,2}$}
\address{$^1$Division of Superconductivity and Magnetism,
Universit\"{a}t Leipzig, Linn\'{e}stra{\ss}e 5, D-04103 Leipzig,
Germany}
\address{$^2$Laboratorio de
F\'isica de Sistemas Peque\~nos y Nanotecnolog\'ia,
 Consejo Superior de Investigaciones Cient\'ificas, E-28006 Madrid, Spain}

\begin{abstract}
Point-contact Andreev reflection measurements  of Co/ and
Cu/tungsten-carbide (WC$_x$) contacts are presented. Metallic thin
films were patterned by e-beam-lithography and lift-off; tungsten
carbide superconducting tips were grown directly on the pre-patterned
samples by decomposition of a metallo-organic vapour (tungsten
hexacarbonyl) under a focused Ga$^+$-ion beam (FIB). Current-voltage
measurements as a function of temperature and magnetic field clearly
showed the signatures of Andreev reflection. The experimental
conductance-voltage curves were analyzed within the
Blonder-Tinkham-Klapwijk theory. The results highlight
the possibilities, advantages and disadvantages of using FIB-produced
amorphous WC$_x$ tips for point-contact spectroscopy in metallic
nanostructures.
\end{abstract}


\submitto{\NT} \maketitle

\section{Introduction}
In recent years research activity has focussed on nano\-technological
applications of superconducting and magnetic materials for quantum
information technology and spintronics \cite{prin,dau}. There are
various aspects of interest that cover both fundamental issues such as the
value of the superconducting gap in tiny superconducting structures,
the behaviour of the electronic density of states at interfaces as well as
applications such as the measurement of the ferromagnetic spin
polarization by Andreev reflection. The measurement and analysis of
Andreev reflection \cite{mazin,and} in superconductor/ferromagnet
(S/F) point contacts has proven to be potentially applicable for the
determination of the spin polarization \cite{sou,upa,ji,nad}, but was
also found to be prone to misinterpretations
\cite{auth2003,branford}. On the other hand, point contact
spectroscopy could help to study systematically the physics of
interfaces between two different materials, an issue of growing
importance nowadays.

In this work the superconducting behaviour, i.e. current-voltage
($I-V$) characteristics and the corresponding differential
conductance $G=\partial I/\partial V$, of $\sim 20~$nm diameter
WC$_x$ electrodes fabricated by a metallo-organic vapour deposition
process was explored. It is shown that Andreev reflection is clearly
observed at WC$_x$ contacts with normal as well as with ferromagnetic
metals. The differential conductance curves, however, show variations
in the local superconducting gap value upon the material on which the
superconductor is deposited. Some of these curves have also an
unconventional form that suggests either the formation of weak links
in the WC$_x$ tip or the contribution of a narrow-band normal metal
at the junction. In spite of its flexibility, it appears that the
necessity of Ga$^+$ion irradiation to produce the superconducting
nanocontact with a focused ion beam (FIB) device influences the
intrinsic properties of the bulk materials at their surfaces.
Nevertheless, this influence should not be taken always as a
disadvantage due to the new physics one may expect from these
materials and at their interfaces. This work opens up the possibility
of growing Andreev reflection test leads on a nanometer scale during
the preparation and handling of different materials in a dual beam
microscope.
\section{Experimental Details \label{experimental}}
%
\begin{figure*}
\begin{center}
\vspace*{0.0cm}
\includegraphics[width=0.7\textwidth]{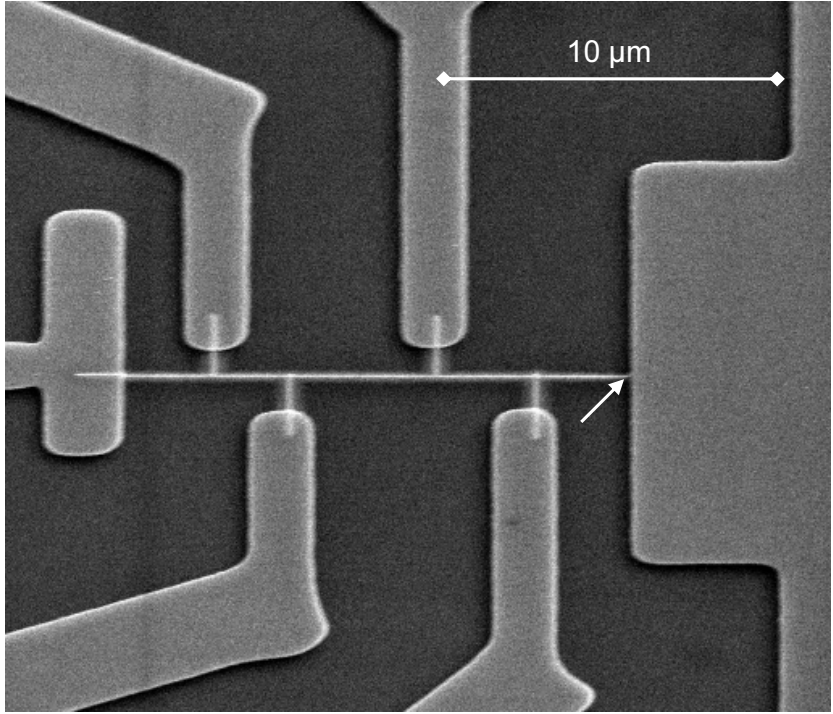}
\vspace*{0.0cm}
\end{center}
\caption{Scanning electron microscopy  image under 52$^\circ$ of a
typical sample used for Andreev-reflection measurements. The arrow
shows the location of the superconducting WC$_x$ wire near the point
contact, see Fig.~\protect\ref{tip}.} \label{sample}
\end{figure*}
\begin{figure*}
\begin{center}
\vspace*{0.0cm}
\includegraphics[width=0.7\textwidth]{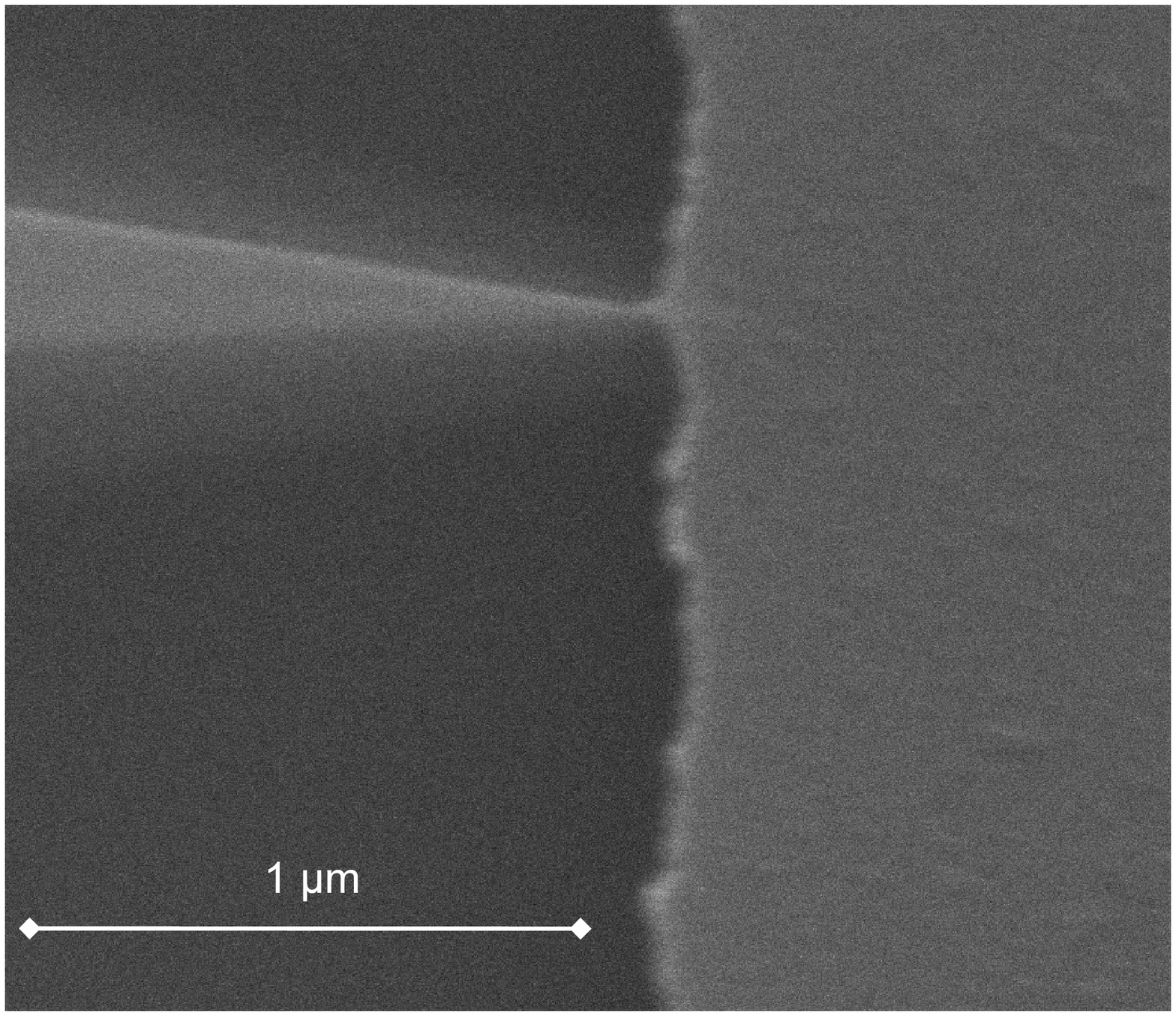}
\vspace*{0.0cm}
\end{center}
\caption{Scanning electron microscopy  image under 52$^\circ$ of the
WC$_x$ point contact on the Co film surface.} \label{tip}
\end{figure*}
For e-beam lithography, sample imaging and metallo-organic vapour
deposition a dual beam microscope (DBM), model FEI NovaLabXT 200,
equipped with the Raith ELPHY Plus electron  lithography system was
used. The sample preparation was realized in two stages. In the first
step the contact layout structure allowing for Andreev reflection
measurements in four-point configuration was fabricated. To this end
Si substrates coated with 150~nm thick SiN films were spin-coated
with a positive resist and the sample layout was transferred to the
resist by e-beam lithography. After developing and cleaning Co or Cu
thin films were sputtered onto the substrates at a base pressure of
$10^{-7}$~mbar using metallic targets of $99.9$\% purity. The
following lift-off concluded the first sample preparation stage with
the production of the contact structure, see Fig.~\ref{sample}.

In the second stage the facilities offered by the DBM were employed,
especially the deposition of metallic films in nanometer dimensions
from the induced decomposition of a chemical precursor over a
substrate by the ion beam. This technique is known as
ion-beam-induced- or -assisted-deposition (IBID or IBAD). The main
advantage of this technique is the deposition of the material in
principally any shape and size without a mask directly onto the
substrate. For the growth of the superconducting WC$_x$ tips tungsten
hexacarbonyl, W(CO)$_6$, was used as a precursor. WC$_x$ deposition
was done under Ga$^+$ ion irradiation with a beam energy of 30~keV
and a focused ion-beam current of 98~pA. The precursor temperature
was set to 56$^\circ$C; during the film deposition the pressure in
the chamber was $\sim 2\times10^{-5}$~mbar. In this way
superconducting nanowires with a cross-section of $250\pm10$\ nm
$\times$ $250\pm10$\ nm were deposited. The end of the nanowire
making the contact to the metal under investigation (see arrow in
Fig.~\ref{sample}) was made in a needle shape such that the contact
area is estimated to be circular with a diameter of about $10\ldots
20$~nm. This tip is shown in Fig.~\ref{tip} for the WC$_x$/Co sample.

The fabrication of the metallic films and nanocontacts was done in
high vacuum which is advantageous over the conventional mechanical
point-contact technique \cite{blon}. Furthermore, several WC$_x$ tips
can be grown on the same substrate such that it would be possible to
study the spatial variation of the properties at the interfaces like
the spin-polarization. The tungsten to carbon ratio in WC$_x$ varies
slightly from sample to sample, but nevertheless the critical
temperatures were consistently around 5~K in agreement with previous
reports \cite{sad,spo}. In spite of these advantages we note that the
IBID process may produce a non-negligible change in the intrinsic
properties of the material surface to be investigated. The changes
depend on the acceleration energy of the Ga$^+$-ions used, the mass
density and other intrinsic properties of the material in question.
We note that the properties of these interfaces were not yet studied
systematically in the past. An anomalous behavior of the differential
conductance, for example, may imply the existence of unconventional
material properties at the interfaces.

After preparation, the samples were mounted in a standard chip
carrier. Contacts were made by Au wires and silver paste.
Current-voltage ($I$--$V$) measurements in four-point configuration
were made with a Keithley DC and AC current source (Keithley 6221)
and a nanovoltmeter (Keithley 2182). The measurements were performed
in a He-flow cryostat (Oxford Instruments) in the temperature range
between 2~K and 300~K with a temperature stabilization better than
1~mK.

\section{Results and Discussion\label{results}}
%
\begin{figure}
\begin{center}
\vspace*{0.0cm}
\includegraphics[width=0.48\textwidth]{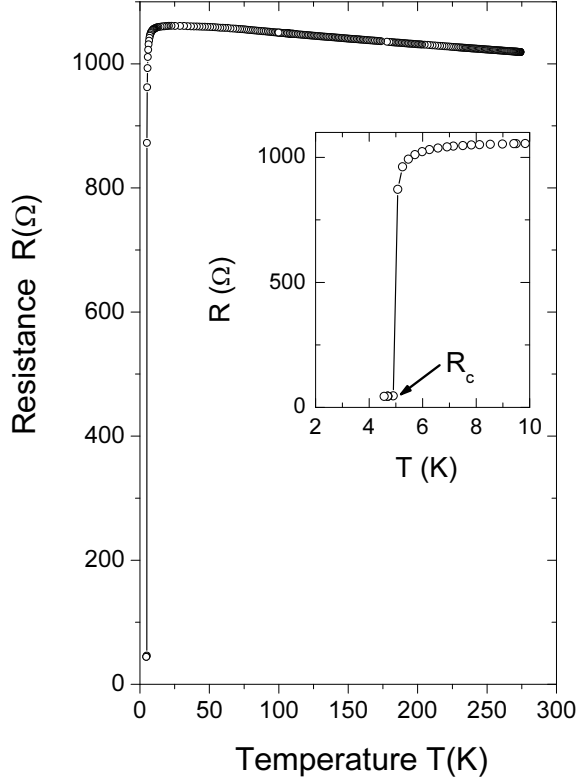}
\vspace*{0.0cm}
\end{center}
\caption{Temperature  dependence of the point contact between WC$_x$
and Co. The inset show a magnified view of the region near the
superconducting transition. For the contacts with Cu a $T_c = 4.10$~K
was measured and a lower residual resistance.} \label{rt}
\end{figure}

Recently published work \cite{sad} reported that WC$_x$ nanowires
produced via the IBAD technique are superconducting; we have reported
\cite{spo} similar results on our WC$_x$ nanowires, where critical
temperatures $T_c \simeq 4 \ldots 5$~K were found. These critical
temperatures are in good agreement with values from amorphous thin
films of W and C produced by RF sputtering \cite{bond}. The upper
critical field was $B_{c2}(T/T_c = 0.75) \simeq 6$~T defined at 90\%
of the normal state resistance \cite{spo}. In situ energy dispersive
X-ray analysis (EDX) was used to study the composition of the WC$_x$
wires and yielded concentrations of about 35\% C, 4\% O, 13\% Ga, 3\%
Se, 28\% Si and 17\% W. The tungsten concentration is in agreement
with the data from Refs.~\cite{spo,bond,koh,jenk}.

Figure~\ref{rt} shows the temperature dependence of a Co/WC$_x$
point-contact resistance. A clear transition in the resistance is
seen when the sample enters into the superconducting state and at 5~K
a constant, residual resistance is reached. This residual resistance is not
zero, but finite with a value of $R_c =47~\Omega$ for this contact
which is just the resistance of the Co metal in the current path.
For the Cu/WC$_x$ junctions we obtained $T_c = 4.1~$K and $R_c =
1~\Omega$. The temperature dependence of the resistance of the
Co/WC$_x$ sample showed a linear behaviour above 45~K with a negative
temperature slope, see Fig.~\ref{rt}, a resistivity ratio $\rho_{50\,
{\rm K}}/\rho_{300\, {\rm K}} = 1.044$ and a normal state resistivity
of about 140~$\mu\Omega$cm at room temperature. A similar dependence
was obtained for the Cu/WC$_x$ samples. More details about the
electrical properties, especially the temperature and magnetic field
dependence of these WC$_x$ wires were reported in \cite{spo}.
Point-contact spectroscopy measurements were performed on several
samples of the same deposited metallic system. Here the results for
representative Cu and Co electrodes are presented.

\begin{figure*}
\begin{center}
\vspace*{0.0cm}
\includegraphics[width=0.8\textwidth]{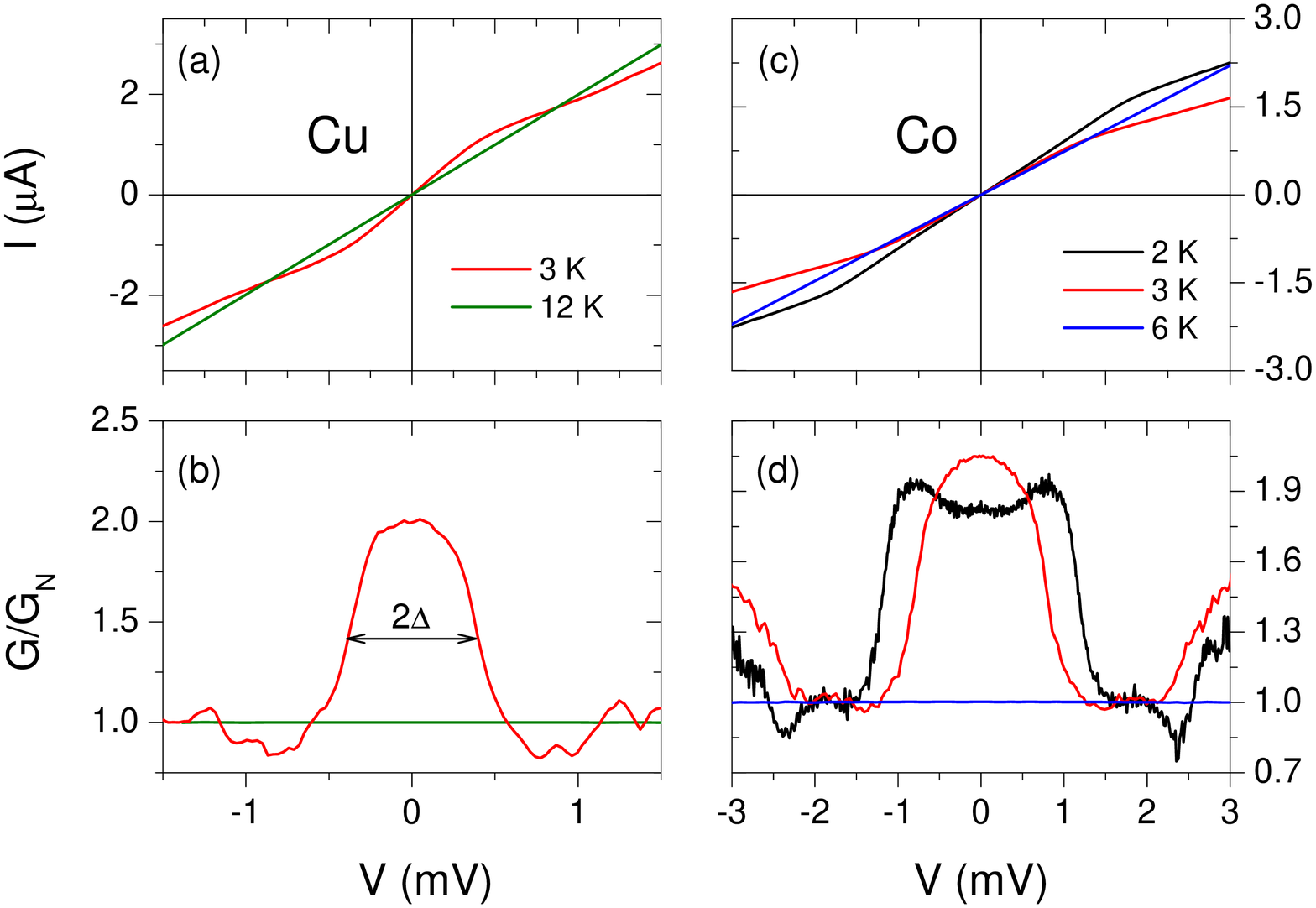}
\vspace*{0.0cm}
\end{center}
\caption{Temperature dependence of the $I$--$V$ curves (a,c) and the
corresponding $G$--$V$ data (b,d) of a WC$_x$/Cu (a,b) and a
WC$_x$/Co (c,d) point contact, respectively.} \label{andreev}
\end{figure*}

Figure~\ref{andreev} presents the $I$--$V$  characteristics of a
WC$_x$/Cu and a WC$_x$/Co point contact measured at various
temperatures. At higher temperatures above the critical temperature
of the WC$_x$ nanowire the $I$--$V$ curves were linear. Below the
critical temperature a clear nonlinearity was observed. This
behaviour is even more obvious in the conductance-voltage
characteristics where the conductance was obtained from the measured
$I$--$V$ curves by numerical differentiation. These curves show that
the WC$_x$ nanocontacts support Andreev reflection.

>From BCS theory the zero temperature gap of the WC$_x$ wires is
estimated to $\Delta(0) = 1.76k_BT_c \simeq 0.76$~meV (Co/WC$_x$) or
$\simeq 0.62~$meV (Cu/WC$_x$). At 3~K the superconducting gap was
determined from the conductance {\em vs.}\ voltage curves as $\sim
0.38$~meV for the Cu/WC$_x$ (in comparison with the BCS value
$\Delta_{\rm BCS}(3~$K$) = 0.49~$meV) and $\sim 0.8$~meV for the
Co/WC$_x$ contact ($\Delta_{\rm BCS}(3~$K$) = 0.68~$meV),
respectively. Whereas for the latter the value of the superconducting
gap is 17\% larger than the expected one from BCS, the value of the
gap for the Cu-based contact is 22\% smaller. These differences are
probably due to interfacial defects.

\begin{figure}
\begin{center}
\vspace*{0.0cm}
\includegraphics[width=0.45\textwidth]{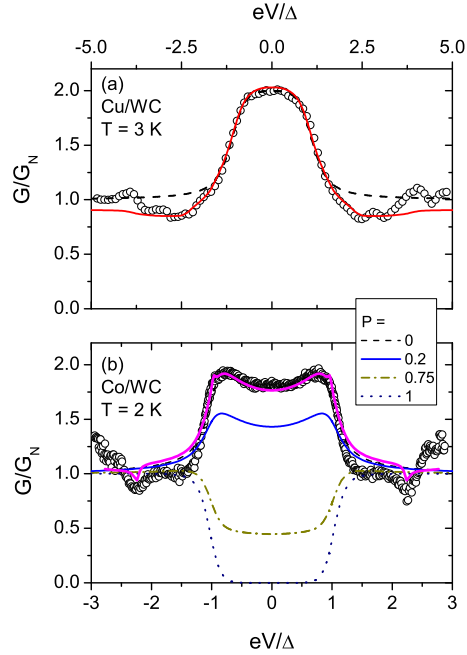}
\vspace*{0.0cm}
\end{center}
\caption{(Color online)(a) Normalized conductance of  a WC$_x$/Cu
point contact at 3~K and zero magnetic field. The dashed line has
been calculated within the BTK model with the parameters $Z = 0$, a
spin polarization $P = 0$ and a gap to temperature ratio
$\Delta/(k_BT) = 6$. The continuous (red) line was obtained adding a
10\% parallel contribution of a narrow band metal with band width $W
= 2.3\Delta$, following \protect\cite{garnw}. (b) The same for the
WC$_x$/Co point contact at 2~K and zero field. The curves were
calculated within the BTK model with $Z = 0.2$, $\Delta/(k_BT) = 12$
and spin polarization values $P$ as indicated in the figure. The
continuous magenta line was obtained adding a 15\% parallel
contribution of a narrow band metal with band width $W = 1.3\Delta$.}
\label{fit}
\end{figure}

In the ideal case, at low temperatures and in the subgap voltage
regime the conductance of a superconductor/normal metal contact
should be twice as large as the normal state conductance. This is
seen in both junctions, see Fig.~\ref{andreev}. However, we observe
clear deviations from the conventional form of the conductance curve
with minima at higher voltages, i.e. at $1.5 \Delta \lesssim eV
\lesssim 3 \Delta$, see Figs.~\ref{andreev} and \ref{fit}. Similar
deviations were reported before
\cite{rowell1968,klapwijk1982,octavio1983,flensberg1988,sou,wei98,sou99,rou05,pan05,str01,panprb05,yun07}
and are characterized by conductance  values less than the normal
state value $G_N$ at $V > \Delta/e$. This behaviour could be
understood either by the formation of a weak link
\cite{klapwijk1982,octavio1983,flensberg1988}  or by the contribution
of a narrow band normal material \cite{garnw}, both might be formed
at the junction interface. Also, it might well be that the weak-link
itself behaves as a narrow band normal material. It can be shown
\cite{garnw} that for a junction between a superconductor with gap
$\Delta$ and a normal metal with a band width $2W$ of the order of
the energy gap, a negative differential conductance appears at
voltages $eV > \Delta$. For example, for transmittivity $T = 1$ and
$W = 2 \Delta$ a clear (negative) minimum appears in $G$ at $eV =
2\Delta$ \cite{garnw}. In all samples the location of the
conductivity minimum appears to scale with the value of the gap. This
may appear unlikely if we assume the existence of this kind of normal
metal independently of the attached superconductor. However, this
narrow band material or the material of the weak link at the
interface may be formed by the contact to the superconducting
material and therefore a direct correlation with the superconducting
energy gap can exist. Phenomenologically speaking, the observed
behavior can be quantitatively understood assuming a parallel
contribution by a material that allows the current to pass through
the junction within an energy band of the order but larger than the
energy gap of the superconducting part.

\begin{figure*}
\begin{center}
\vspace*{0.0cm}
\includegraphics[width=0.8\textwidth]{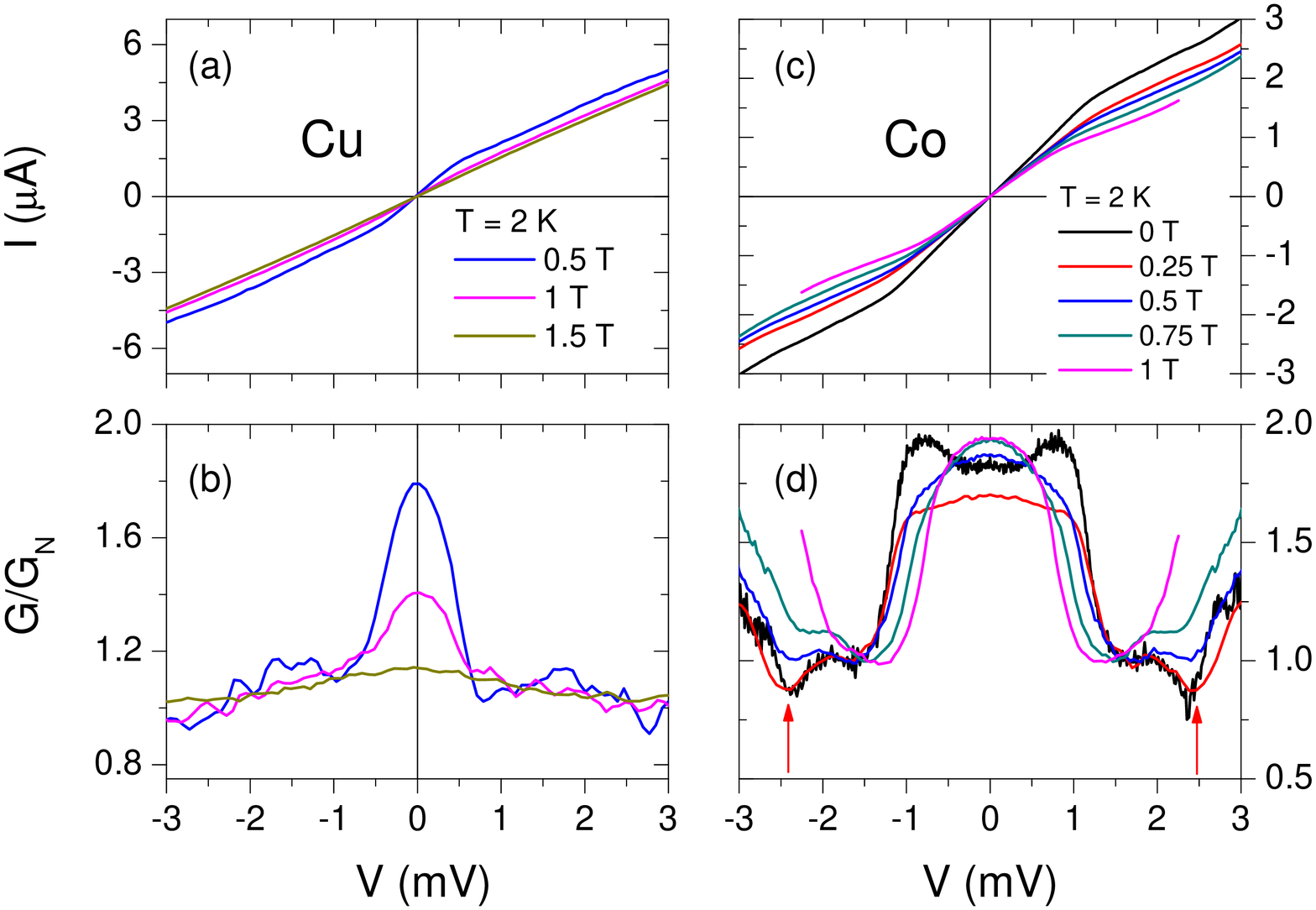}
\vspace*{0.0cm}
\end{center}
\caption{Magnetic field dependence of the  $I$--$V$ curves (a,c) and
the corresponding $G$--$V$ data (b,d) of a WC$_x$/Cu (a,b) and a
WC$_x$/Co (c,d) point contact, respectively.} \label{field}
\end{figure*}
The magnetic field dependence at constant temperature was also
investigated with the results for the two junctions shown in
Fig.~\ref{field}. The magnetic field reduces the zero bias
conductance enhancement and smears out the bias dependent features.
Note that the conductance minimum in Fig.~\ref{field}(d), marked by
the red arrows, is systematically removed by the magnetic field. This
indicates that these features are actually linked to the
superconducting state. Similar magnetic field effects were observed
in experiments on MgB$_2$ \cite{sza} and were explained as a
consequence of the pair-breaking effect; note, however, that MgB$_2$
is a standard example of a superconductor with two intrinsic energy
gap values. The field dependence observed here corroborates the
interpretation of the conductance-voltage curves as arising from
Andreev reflection.

The experimental data were analyzed within the
Blonder-Tinkham-Klapwijk (BTK) model \cite{blon1}
and the narrow band model \cite{garnw}. In the BTK model the quality of the
interface is characterized by a dimensionless parameter for the
barrier strength, conventionally denoted $Z$, with $Z = 0$
corresponding to a clean interface and $Z \gg 1$ corresponding to the
presence of a tunnelling barrier between superconductor and metal.
The transmittivity used in \cite{garnw} can be expressed as $T = 1/(1
+ Z^2)$, where $Z$ was defined in \cite{blon1}. Notice that $Z = 0$
and $\infty$ means $T = 1$ and 0.

From the result $G/G_N \simeq 2$ for both junctions one might
tentatively conclude that $Z \ll 1$, i.e.~that the barrier is
comparatively clean.  Figure~\ref{fit} shows conductance data in
comparison to calculations within the BTK model. Reasonable agreement
was found for subgap voltages.  The parameters used in the BTK
calculations were $Z = 0$, a spin polarization $P = 0$ and a gap to
temperature ratio $\Delta(3$K$)/k_BT = 6$ for the Cu/WC$_x$ junction
as well as $Z = 0.2$ and $\Delta(2$K)$/T = 12$ for the Co/WC$_x$
junction. The gap to temperature ratios $\Delta/k_BT$ needed to fit
the data using the BKT formulas are significantly larger than the
ones determined independently from the measured temperature and gap
values as we showed above. In other words the $G/G_N$ curves at the
measured temperatures for both cases are significantly sharper than
expected from the BKT calculations, a fact that may indicate a
non-BCS temperature behaviour very probably related to the
interfacial properties. Also the vanishing spin polarization of the
Co/WC$_x$ contact can be  due to radiation defects at the Co surface
introduced in the fabrication process. It might as well be related to
the inferior crystallographic quality of our sputter-deposited Co
films.

The conductance minima at larger voltages as well as the increase in
$G/G_N$ at voltages above the minimum at $eV/\Delta \simeq 2.3$, see
Fig.~\ref{fit}(b), cannot be reproduced within the BTK model of a
normal/superconducting contact. Assuming that at  the interface of
the junction a weak link of a material that behaves as having a band
width $W = 2.3 \Delta$ contributing in parallel to the measured
current, one can reasonably well reproduce the observed minimum due
to the negative differential conductivity contribution, see
Fig.~\ref{fit}(a). In a similar way one can understand the observed
minimum in the Co/WC$_x$ junction, decreasing in this case the
absolute band width of the weak link material, see Fig.~\ref{fit}(b).
The observed conductance increase at voltages $eV > 2\Delta$ might be
due to nonlinear contributions to the tunnelling conductance as
discussed in the Simmons's model \cite{simmons1963}.

\section{Conclusions \label{discussion}}
In this work it was shown that Andreev reflection  can be observed in
point contacts between Cu and Co films and WC$_x$ tips deposited by
metallo-organic vapour deposition in a dual beam microscope.
Qualitatively the measured temperature and magnetic field dependence
of the differential conductivity agree with the expectations. The Co
films investigated here did not appear to be spin-polarized at the
contact region according to the conventional BTK model for a
ferromagnet/superconductor contact. On the other hand the parameters
needed to fit the experimental curves at relatively high temperature
appear to be inconsistent with the BCS expectations. The anomalies
observed in the differential conductivity at voltages above the
energy gap values suggest  a weak link formation  with narrow band
properties contributing in parallel at the interfaces of the
contacts. In spite of the disadvantages Ga irradiation may have when
used to deposit the WC$_x$ amorphous superconductor, the presented
method holds future potential in the preparation of Andreev contacts
for spin-polarization measurements on the several nanometer scale,
offering fast and reliable sample preparation and the possibility to
determine the spatial variation of the spin-polarization. Systematic
measurements using different materials can be used in the future to
study in more detail the properties of interfaces, which can be
modified using FIB within a dual beam microscope.

\ack This work was supported by the DFG under DFG ES 86/16-1,  HBFG
grant No.~036-371 and by SFB~762 ``Funkionalit\"at Oxidischer
Grenzfl\"achen". The collaboration between Madrid and Leipzig was
supported by the DAAD under Grant No.~D/07/13369 (``Acciones
Integradas Hispano-Alemanas").

\section*{References}
\providecommand{\newblock}{}

\end{document}